\documentclass{svproc}
\usepackage{url}

\usepackage[pdftex]{graphicx}

\begin{document}
\mainmatter              
\title{\textit{DeepObfusCode}: Source Code Obfuscation Through Sequence-to-Sequence Networks}
\titlerunning{DeepObfusCode}  
\author{Siddhartha Datta \footnote{Work performed at the Hong Kong University of Science and Technology.}}
\authorrunning{Siddhartha Datta et al.} 
%
\tocauthor{Siddhartha Datta}

\institute{University of Oxford,\\
\email{siddhartha.datta@cs.ox.ac.uk}\\ }

\maketitle              

\begin{abstract}
The paper explores a novel methodology in source code obfuscation through the application of text-based recurrent neural network network (RNN) encoder-decoder models in ciphertext generation and key generation. Sequence-to-sequence models are incorporated into the model architecture to generate obfuscated code, generate the deobfuscation key, and live execution. Quantitative benchmark comparison to existing obfuscation methods indicate significant improvement in stealth and execution cost for the proposed solution, and experiments regarding the model's properties yield positive results regarding its character variation, dissimilarity to the original codebase, and consistent length of obfuscated code.
\end{abstract}

\section{Introduction}
The field of code obfuscation has aimed to tackle reverse-engineering of code bases for years. The entire basis of this methodology is that if a program is constructed with logic not easily recognizable by a reader, the logic would be preserved intact and the software would be intrinsically protected. Traditional tactics include creative uses of whitespace, redundant logical operations, unnecessary conditional operations, amongst others. The common issue with obfuscation is that it can be reverse-engineered, the only factor for a malicious actor would be the amount of time needed to discover the logic. \textit{DeepObfusCode} is a proposed methodology to use neural networks to convert the plaintext source code into a cipher text by using the propagating architecture of neural networks to compound the randomness factor in the creation of the ciphertext. Yet at the same time, neural networks have the ability to learn statistical patterns and generate weights to convert one text to another, in our case from the ciphertext to plaintext. This would eventually permit users to simply load the ciphertext and the key to self-execute the program without foreign users viewing the inner workings of the code. From an academic standpoint, this methodology redirects obfuscation methodology towards complete obfuscation in contrary of incremental obfuscation, and suggests the usage and development of asymmetric key infrastructure in obfuscation. Beyond sequence-to-sequence network models, further obfuscation models could be built with greater degree of resilience, and other deep learning methods could be harnessed to develop alternative techniques to obfuscate code. The methodology can be adopted for more efficient, more effective obfuscation for developers protecting their proprietary codebase or cloud computing services wishing to guarantee confidentiality to customers. The algorithmic architecture could be further incorporated into larger frameworks or infrastructures to render homomorphic encryption and ensure complete anonymity of the codebase during execution from the execution provider.

\section{Related Work}
This section will detail existing code obfuscation methods being used, how they are evaluated, and how modern deep learning techniques are being used in the area, while drawing comparison to the proposed obfuscation implementation. 

\subsection{Code Obfuscation Methods}
The objective of code obfuscation is to mask the computational processes and logic behind software code, to protect trade secrets, intellectual property, or confidential data. As such, there have been eight generalized methods behind code obfuscation [1], namely: (i) Name obfuscation; (ii) Data obfuscation; (iii) Code flow obfuscation; (iv) Incremental obfuscation; (v) Intermediate code optimization; (vi) Debug information obfuscation; (vi) Watermarking; (vii) Source code obfuscation. Source code obfuscation, the focus of the paper, is the process that hides the meaning behind the source code, in the case that a third party obtains the code that the code renders itself un-understandable. Under each branch of obfuscation method, there are sub-techniques to reduce understandability of code that are shared by the other obfuscation methods, including control ordering (changing the execution order of program statements), control computation (changing the control flow of the program, such as inserting dead code to increase code complexity), data aggregation (changing the access to data structures after their conversion into other types), renamed identifiers (replacing identifiers with meaningless strings as new identifiers to reduce readability and comprehension of what certain functions or methods do). While existing methods tend to require manual altering of source code with the aforementioned methods, the proposed method performs complete obfuscation with relatively randomly generated characters as the code output. Malicious attackers or readers of the code will not be able to reverse-engineer the code based on the readability of the obfuscated code, and a well-lodged key file of the model weights (e.g. kept on the execution server) would prohibit de-obfuscation.

Existing methods of evaluating obfuscation techniques have tended towards qualitative surveys of difficulty and time taken to de-obfuscate by students, software engineers or computer scientists [2]. To quantitatively compare the performance of the proposed source code obfuscation method against existing code obfuscation methods, we will modify a framework that has been used to compare obfuscation methods before [3], but also malleable enough for us to adapt for our specific comparison use case. The four original comparison metrics are: 

\textbf{(i) Code potency}: This metric arbitrarily computes the degree of obfuscation with traditional complexity measures, with specific focus on control flow and data obfuscation. An example of an implementation would be frequency counts of misleading statements.

\textbf{(ii) Resilience}: This metric measures the ability of the obfuscated text to withstand attacks from automated tools.

\textbf{(iii) Stealth}: This metric tests the difficulty in manual de-obfuscation by humans. It inherently checks how quickly adversarial de-obfuscators can detect misdirecting statements, correctly interpret supposedly-confusing identifiers, and tests their ability to uncover the logic and process behind the code without prior or minimal knowledge of the program.

\textbf{(iv) Execution cost}: This metric measures (i) the incremental time required to perform the obfuscation, and (ii) the incremental time required to execute the obfuscated code. This would be contrasted to the time taken for the original code without obfuscation. 

Since the proposed obfuscation method generates a ciphertext completely different from the original source text and cannot be reverse-engineered by manual human de-obfuscation, the former two metrics (code potency and resilience) would not be used as comparison metrics for this method. The main metrics for evaluating \textit{DeepObfusCode} would be stealth and execution time.

\subsection{Applications of Neural Networks}
Neural networks have had recent applications in the field of cryptography and stenography. There has been a recent implementation of style transfer in stenography [4], where secret images are encrypted within a cover image, and the secret could be obtained by passing the encrypted image through a secondary network. There is another implementation of n-dimensional convolutional neural networks used to encrypt input data [5]. Another implementation involves the use of a multilayer perception (MLP) model to encrypt satellite images, and decrypting using a secondary MLP model [6]. Implementations on neural networks that can execute on encrpyted data has also been a recent development [7]. This indicates a growing interest in encrypting data through the use of neural networks. Unlike prior on data encryption through deep learning techniques, the proposed architecture opens an alternative application, which is to encrypt source code itself through the use of deep learning, then using the generated model file to decrypt and execute the code. 

RNN Encoder-Decoders or sequence-to-sequence models, trained to predict an output text set from provided input text, have had growing applications in statistical machine translation [8], grammatical error correction [9], and text summarization [10]. The notion of taking input sentences and converting them to labeled output sentences with a trained model of weights for character-by-character prediction would be the basis for the proposed obfuscation method. The proposed obfuscation utilizes the advantage of text-based encoder-decoders in character generation and calculating weights to convert one string into another.

\section{\textit{DeepObfusCode}}
The architecture behind the obfuscation is first a primary recurrent neural network (RNN) encoder-decoder model with randomly-set weights taking the original code text as an input to generate the obfuscated text. Then a secondary RNN encoder-decoder model is passed two arguments, the generated ciphertext and the original code as the corresponding label, and is trained for a number of iterations to calculate weights to generate the original text, and the weights of the network would be the key generated. This section will discuss the architecture and implementation in further detail. Supporting code is available \footnote{Code repository: \url{https://github.com/dattasiddhartha/DeepObfusCode}}.

\subsection{Ciphertext generation}

\begin{figure}[h!]
  \centering
  \includegraphics[scale=0.35]{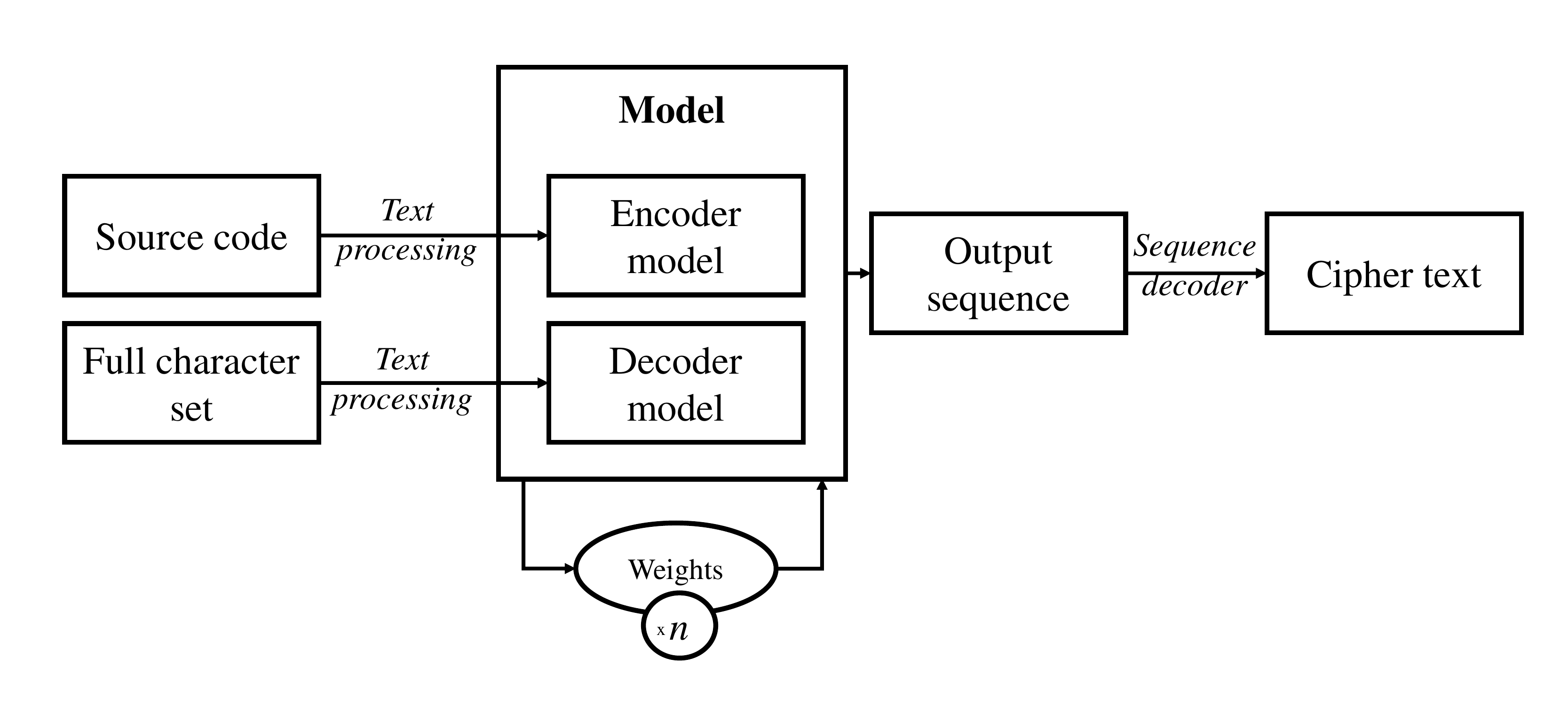}
  \caption{Overview of ciphertext generation. We pass the source code and a character set as inputs to initialize both character sets for the encoder-decoder model, then randomly assign weights to generate the ciphertext, an obfuscated version of the source code.}
\end{figure}

To generate the obfuscated code, we first take the original legible text (source code) and a full character set (a string containing all characters, including letters, numbers and punctuation) (Figure 1). 

We perform text processing for both text strings to prepare the inputs for the encoder and decoder models. First we create unique character sets for both strings, then create two dictionaries for both strings to index each character (key is index and value is character), and create two other dictionaries for both strings to return character given its index (key is character and value is index). Finally we vectorize both strings. 

We use the source code character set as inputs for the encoder model and both text character sets for the decoder model. We will randomly generate weights for a given number of times \(n\) (the randomness index; in our experiment we set \(n=10\)), and set the model weights to be the generated weights array. The variation in the weights array will determine the degree of randomness in the ciphertext.

Weights generated will alter the character value generated for each segment of the string, as shown in the Ciphertext Generation function \(C(p)\). \(c\) refers to the ciphertext (obfuscated code), \(p\) refers to the plaintext (source code), \(N\) refers to each weight position. \(f_n\) and \(w_{rand}\) are the \(n\)-th feature and randomized weight, respectively. \(Z(p)\) is a normalization constant that is independent of the weights.

\[ C(p) = \log p(c|p) = \sum_{n=1}^{N} w_{rand} f_n(c|p) + \log Z(p)  \]

The output will be the output sequence, to which we will decode by taking reference from the index-to-character dictionary to convert the output sequence into an output character string and iteratively append characters. The resulting output would be the ciphertext.

\subsection{Key generation}

\begin{figure}[h!]
  \centering
  \includegraphics[scale=0.35]{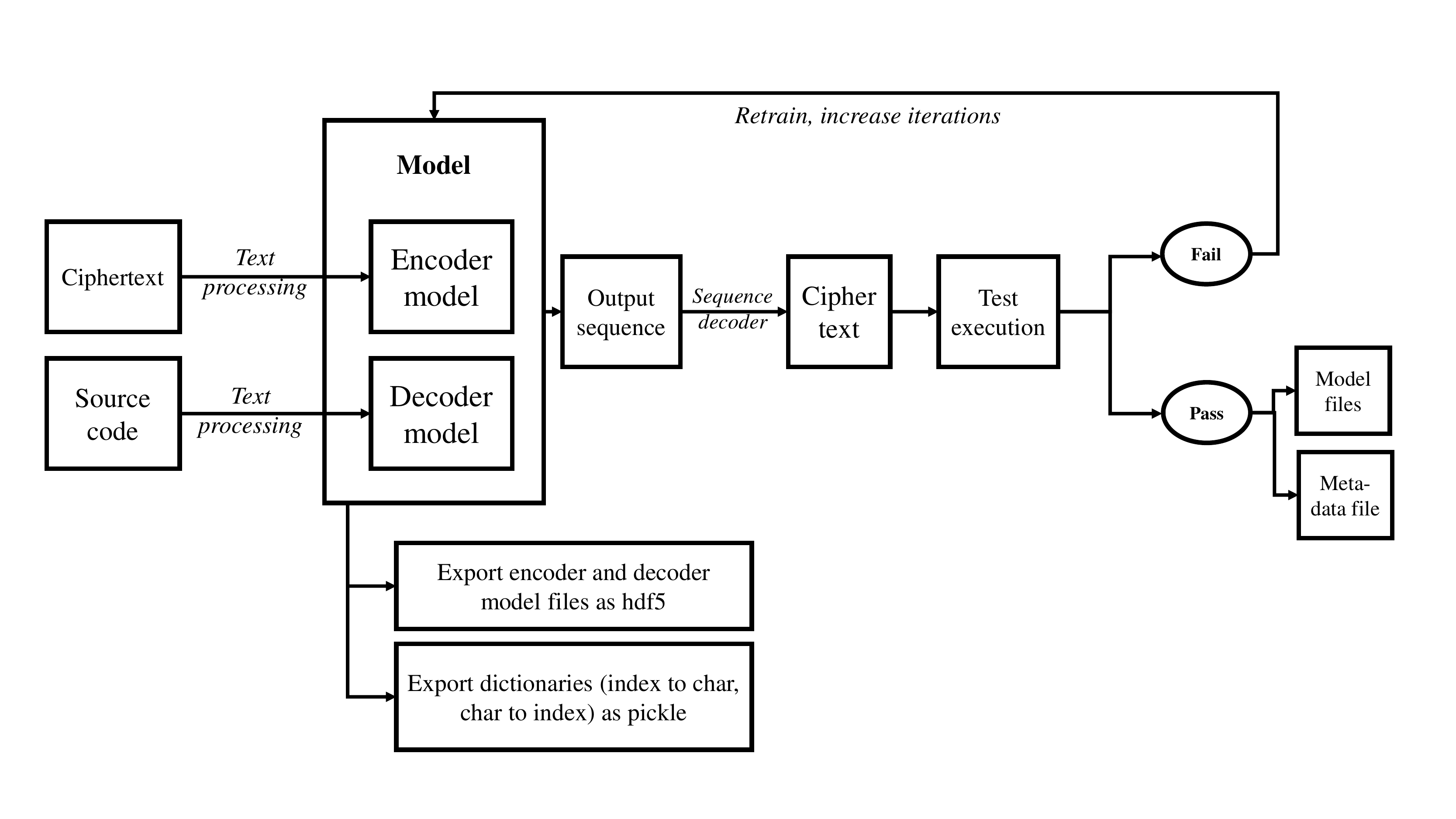}
  \caption{Overview of key generation. With the known ciphertext and original source code, the developer of the source code would pass them as inputs into another encoder-decoder model and train over a number of iterations such that the model weights obtained can translate the obfuscated code into executable code, with validation of executability at the end.}
\end{figure}

After generating the ciphertext, we use it along with the original source code plaintext to generate a key (Figure 2). We apply the same text processing steps as before, set ciphertext as input into the encoder, and both texts as inputs into the decoder, and train the model for a certain number of iteration counts. For our experiments, we tended to use 2000 iterations, though Early Stopping mechanisms could be used to lower training time for shorter source code; the iteration count should be the number of iterations needed to ensure the output text is executable and identical to the source code text. After the training and model weight setting process is complete, export the encoder and decoder model files in HDF5 format (the key), and export the metadata (the dictionaries of index to char, and char to index) in pickle format. The Key Generation function \(K(p,c)\) accepts the arguments \(c\) ciphertext and \(p\) plaintext, and sets weights while minimizing the loss function. \(f_n\) and \(w_n\) are the \(n\)-th feature and weight (after training), respectively. \(Z(c)\) is a normalization constant that is independent of the weights.

\[ \max \frac{1}{N} \sum_{n=1}^{N} \log p(p|c) \]
\[ K(p,c) = \log p(p|c) = \sum_{n=1}^{N} w_n f_n(p|c) + \log Z(c)  \]

After the ciphertext is decoded from the output sequence, we test if the code is executable; if it fails, we retrain the model (a pre-determined loss threshold that ensures correct execution would facilitate early stopping in iterative training); if it passes, the ciphertext and key pair are retained.

\subsection{Source code execution}

\begin{figure}[h!]
  \centering
  \includegraphics[scale=0.35]{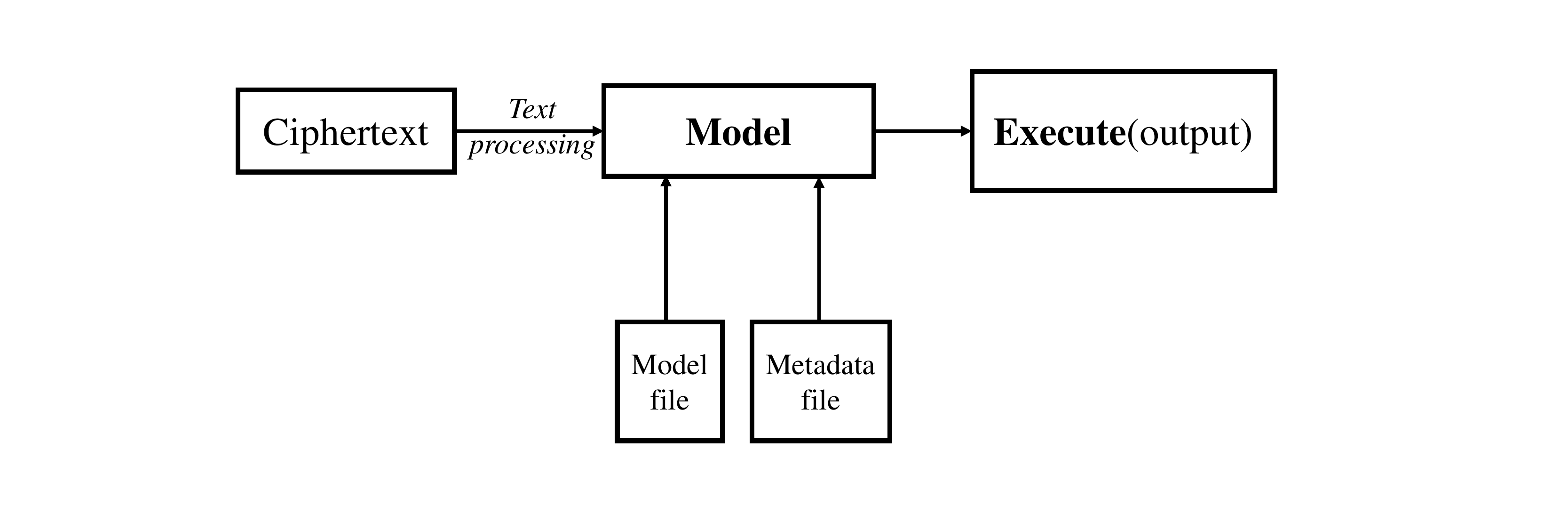}
  \caption{Overview of live execution. To run the obfuscated code on any server or system, one would pass in the obfuscated code into an execution engine that takes the ciphertext and the lodged model files as inputs to execute the withheld code.}
\end{figure}

During live execution (Figure 3), we have three inputs: the obfuscated code (ciphertext), the key (model files), and the metadata files; function \(K(c, k)\) would be modified to accept \(c\) ciphertext, and \(k\) model file (along with metadata file). For our own experiments, the model and metadata files were separate; for execution in live systems, it would be possible to combine them into a single file if preferred. When we pass all three through the model container, the output value is executed as soon as it is returned, i.e. \( Exec (K(c, k))\).

\section{Results}
The evaluate the obfuscation method compared to existing obfuscation techniques, the method will be tested on two aspects, stealth and execution cost. Other parameters such as code potency or resilience would not be applicable to this method, as those comparison metrics require some form of the original code to be preserved; but since this obfuscation method completely regenerates the code base to be indistinguishable from the original code, such testing (e.g. searching for misleading control flow) will not work.

\subsection{Stealth}
The objective of testing for the stealthiness of obfuscated code is to measure the complexity or distance in randomness from the original code. Hence the logic behind our comparison is to first obtain recognized well-obfuscated code and its deobfuscated copy (the original code), obfuscate the original code using our proposed method, then compare the distance in similarity between the original code and each of the obfuscated code. 

Obfuscated code samples were obtained through the International Obfuscated C Code Contest (IOCCC), and de-obfuscated code samples of the selected obfuscated code was found on forums. The obfuscated samples were winners in the IOCCC contest, signifying high quality, and were manually created by programmers. A total of five obfuscated-deobfuscated paired samples were used for the experiment\footnote{Dataset: \url{https://github.com/dattasiddhartha-1/obfuscation-dataset}}. Obfuscated code of reputable, unbiased quality and its deobfuscated counterpart is generally difficult to obtain in large samples; the aforementioned samples would be used in bench-marking and comparison, but not for testing the inherent properties of the proposed architecture. 

To test the obfuscation method extensively, for each de-obfuscated sample, we performed the obfuscation to obtain the ciphertext for 100 iterations (to return 500 samples of obfuscated code samples in total, 100 per de-obfuscated code sample). 

Then we used Levenshtein Distance to measure the distance between two code samples; the distance is the number of deletions, insertions, or substitutions required to transform the source string into the target string. After calculating the Levenshtein Distance values between the original de-obfuscated code and the IOCCC obfuscated code, we calculated the Levenshtein Distance values between the original de-obfuscated code with each respective sample's \textit{DeepObfusCode} obfuscated code and took the mean Levenshtein Distance for the sample. The results are tabulated in Table 1. 

\begin{table}
\renewcommand{\arraystretch}{1.3}
\caption{Levenstein distance comparison}
\label{table_example}
\centering
\begin{tabular}{p{2.2cm}||p{2.4cm}||p{2.2cm}||p{3cm}}
\hline
\bfseries Deobfuscated Code Set & \bfseries Benchmark Obfuscation Lev. Distance & \bfseries Proposed Obfuscation Lev. Distance & \bfseries Ratio of Proposed : Benchmark\\
\hline\hline
1 & 3061 & 4000.68687 & 1.30699\\
2 & 2587 & 497.17172 & 0.19218\\
3 & 42 & 102.81818 & 2.44805\\
4 & 4649 & 5780.32323\ & 1.24335\\
5 & 9132 & 10195.40404 & 1.11645\\
\hline
\end{tabular}
\end{table}

From the table, we can observe that in general the \textit{DeepObfusCode} obfuscation has a greater degree of randomness or dissimilarity from the original source text compared to the benchmark IOCCC obfuscated text, with an average magnitude improvement of  1.2614 (the average of the ratio of proposed:benchmark across all samples) and with a standard deviation 0.7176. 

Set 2 of the experiment indicates the benchmark outperforming the proposed method. Inspection of set 2 reveals a greater proportion of repetitive characters as junk code insertions, which serve to confuse de-obfuscators, but also dilute the similarity (conversely inflate the Levenstein distance) since the proportion of overlapping characters to the total number of characters would be lower. While the total collection indicates a general improvement in dissimilarity, removing set 2 would indicate an average magnitude improvement of 1.5287 and a standard deviation of 0.5352. This aids in justifying that the proposed obfuscation method performs well in the aspect of stealth.


\subsection{Execution Cost}	
The primary focus of the section would be to measure how ciphertext generation time (time to encrypt) and key generation time (time to decrypt) would vary with the length of the source code (plaintext length). 

\begin{figure}[h!]
  \centering
  \includegraphics[scale=0.395]{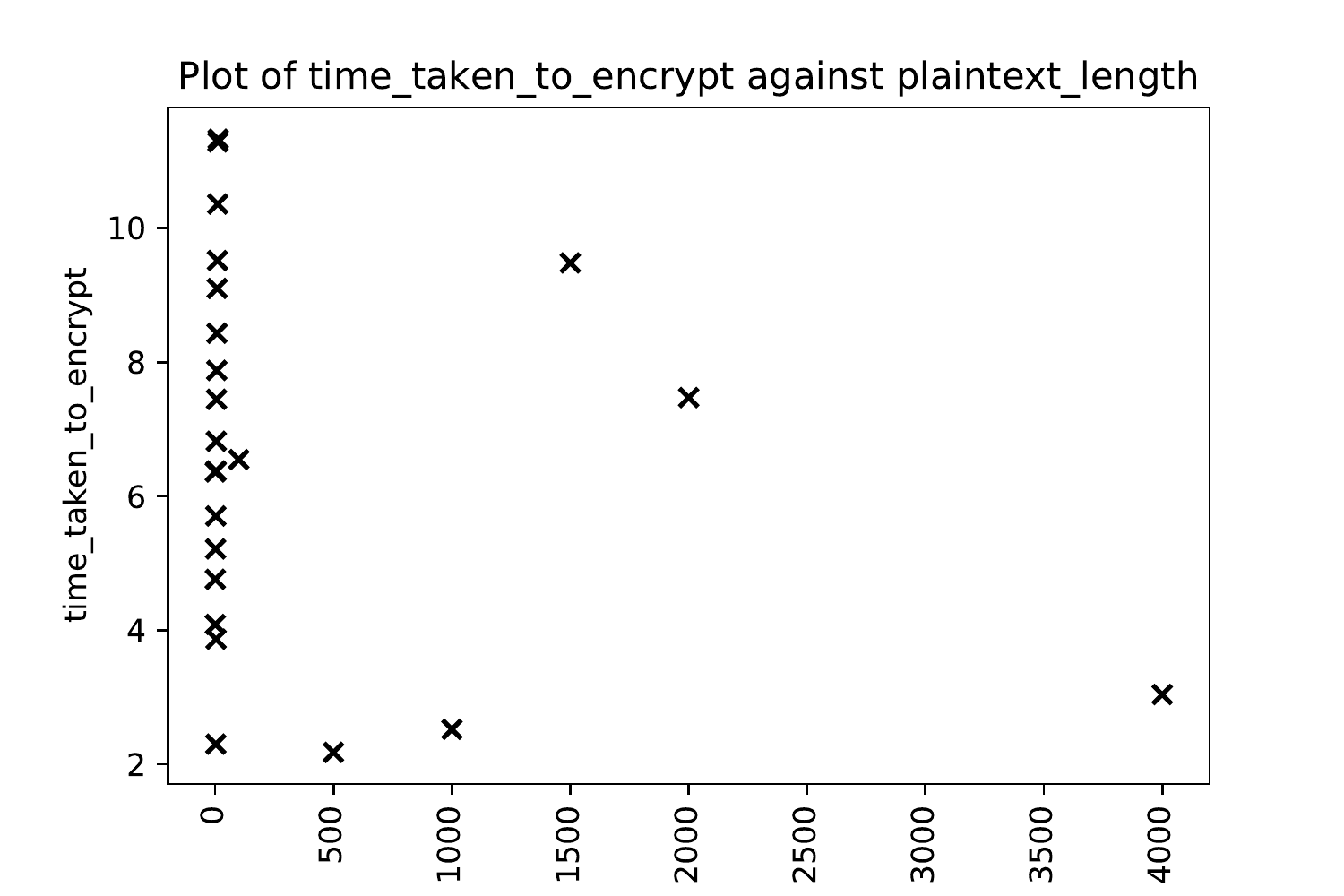}
  \includegraphics[scale=0.395]{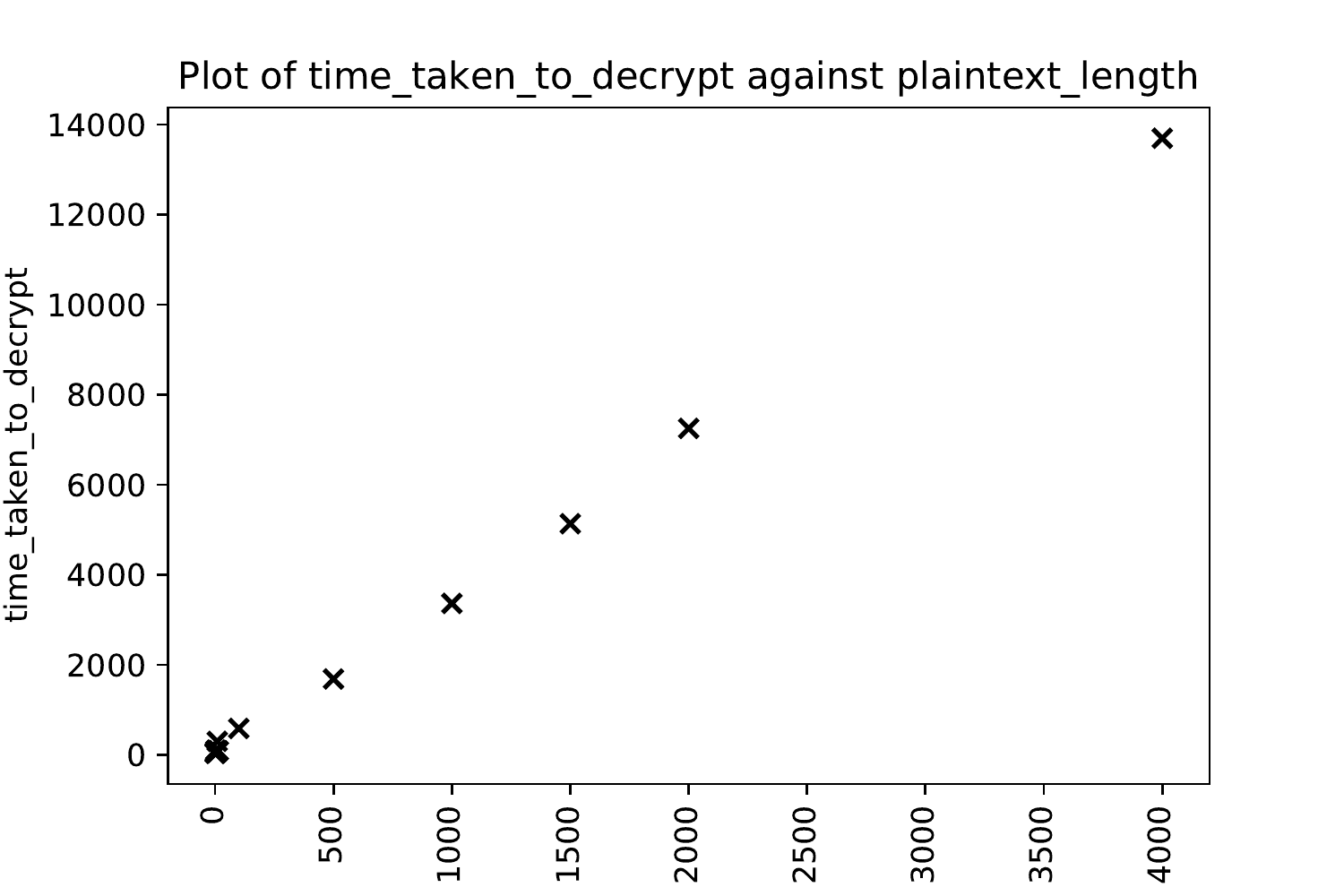}
  \caption{Plots of time properties against plaintext length}
  \centering
  \includegraphics[scale=0.395]{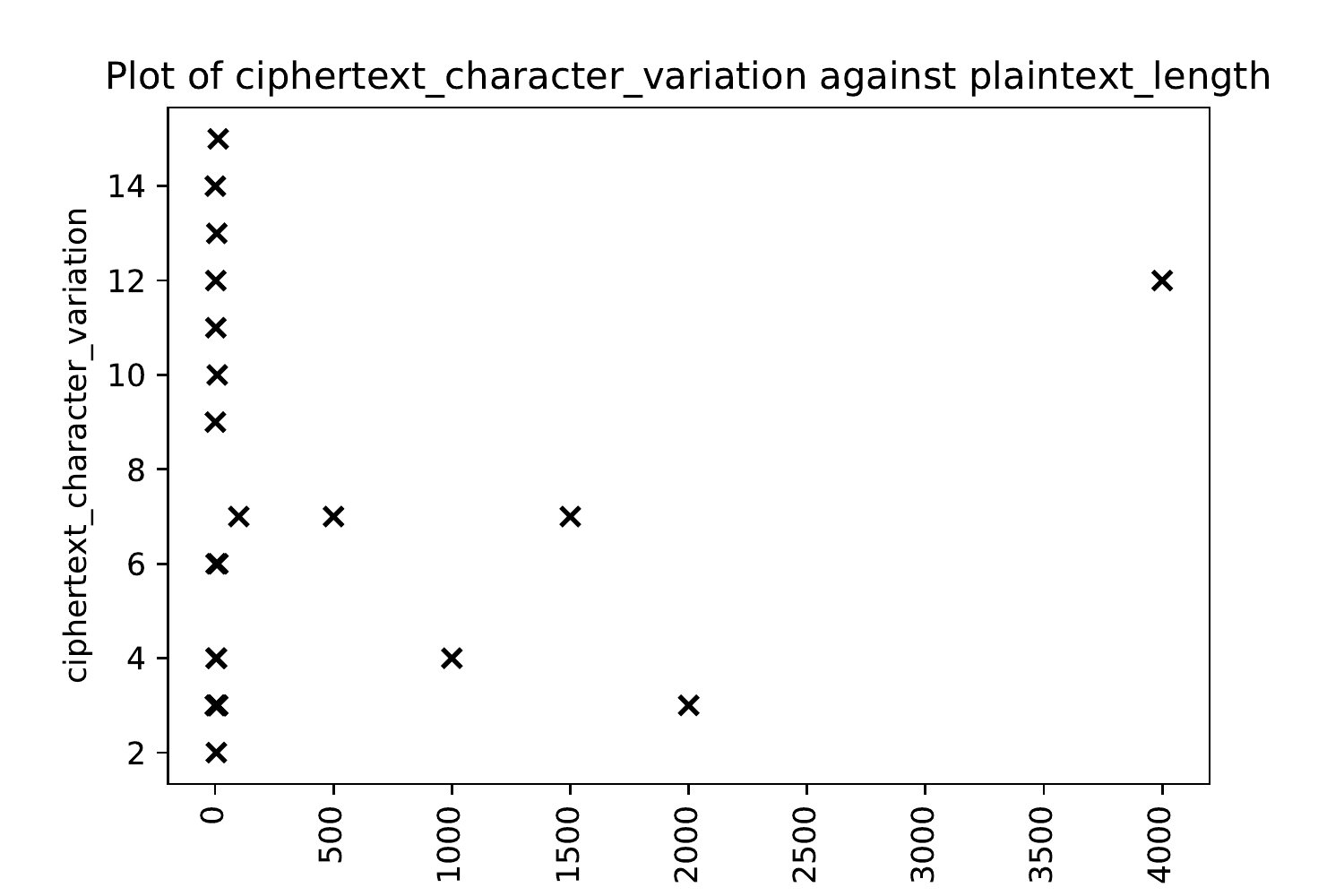}
  \includegraphics[scale=0.395]{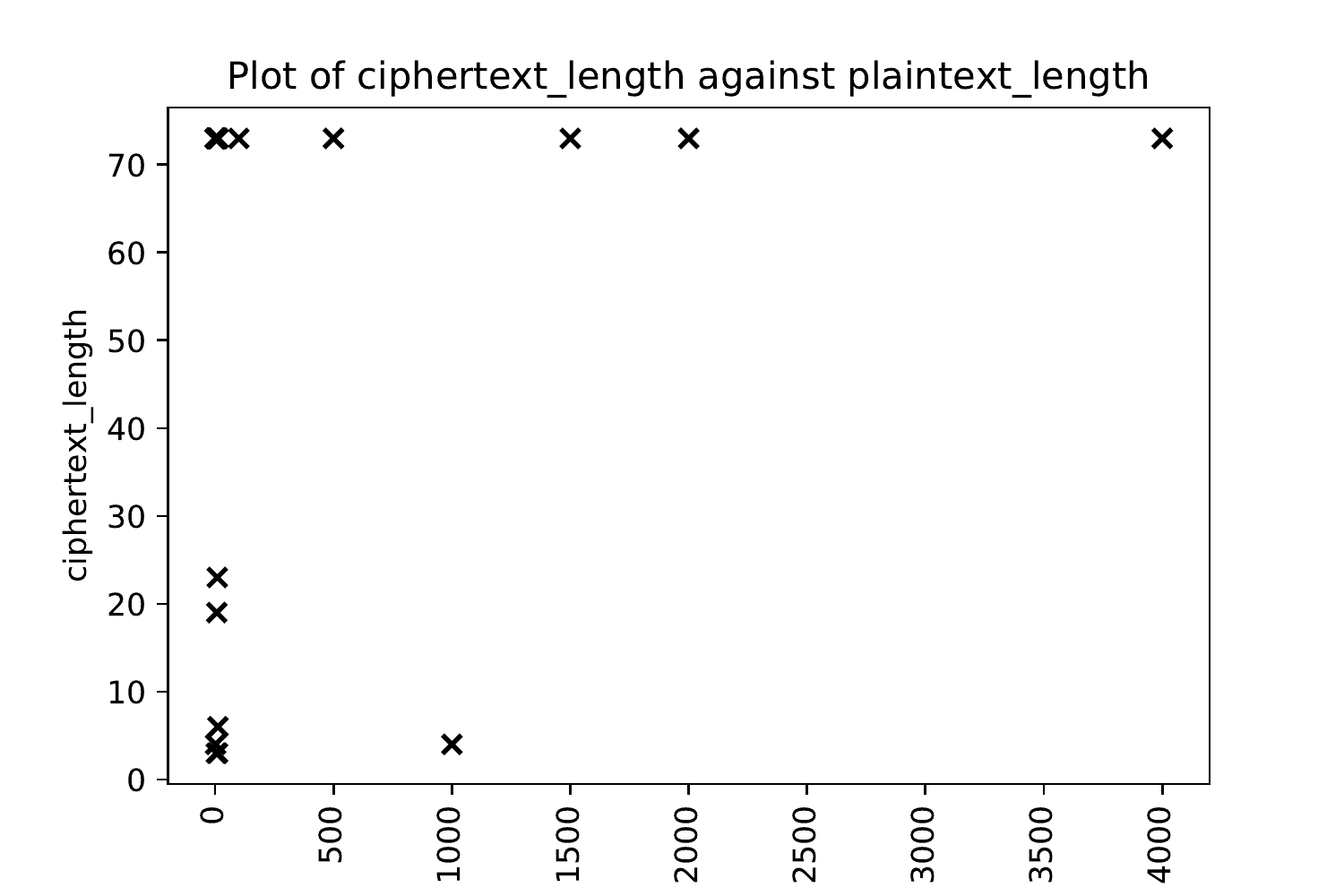}
  \caption{Plots of ciphertext properties against plaintext length}
\end{figure}

The experimental design starts with the random generation of strings varying from length 1 to 4000 (to vary the plaintext length). Then we set loggers to record information regarding the six main variables we would like to test: (i) Length of string input, (ii) Randomness metric (Levenshtein distance), (iii) Execution time for encryption, (iv) Execution time for decryption, (v) Character variation, (vi) Average character length of ciphertext. As execution time depends on the device running the simulation, for reference, the simulation was run on a Python Jupyter notebook, running on Windows 10 with a Nvidia Geforce 1070 GTX graphics card (running at 10\% GPU usage level). 

The plot of ciphertext generation time requirement against length of the source code shows no distinct pattern, while the plot of key generation time requirement against length of the source code shows a linear pattern. This infers that ciphertext generation is not length-dependent and can be executed without significant incremental cost. Since the length of the source code affects the training time required for the same number of iterations (higher length would increase training time per epoch), longer source code would require more time to generate a key, so this obfuscation method may be more suitable for smaller code bases or systems with sufficient computing resources. We can also infer the key generation takes linear time, i.e. time complexity is \(O(n)\). 

\begin{figure}[h!]
  \centering
  \includegraphics[scale=0.425]{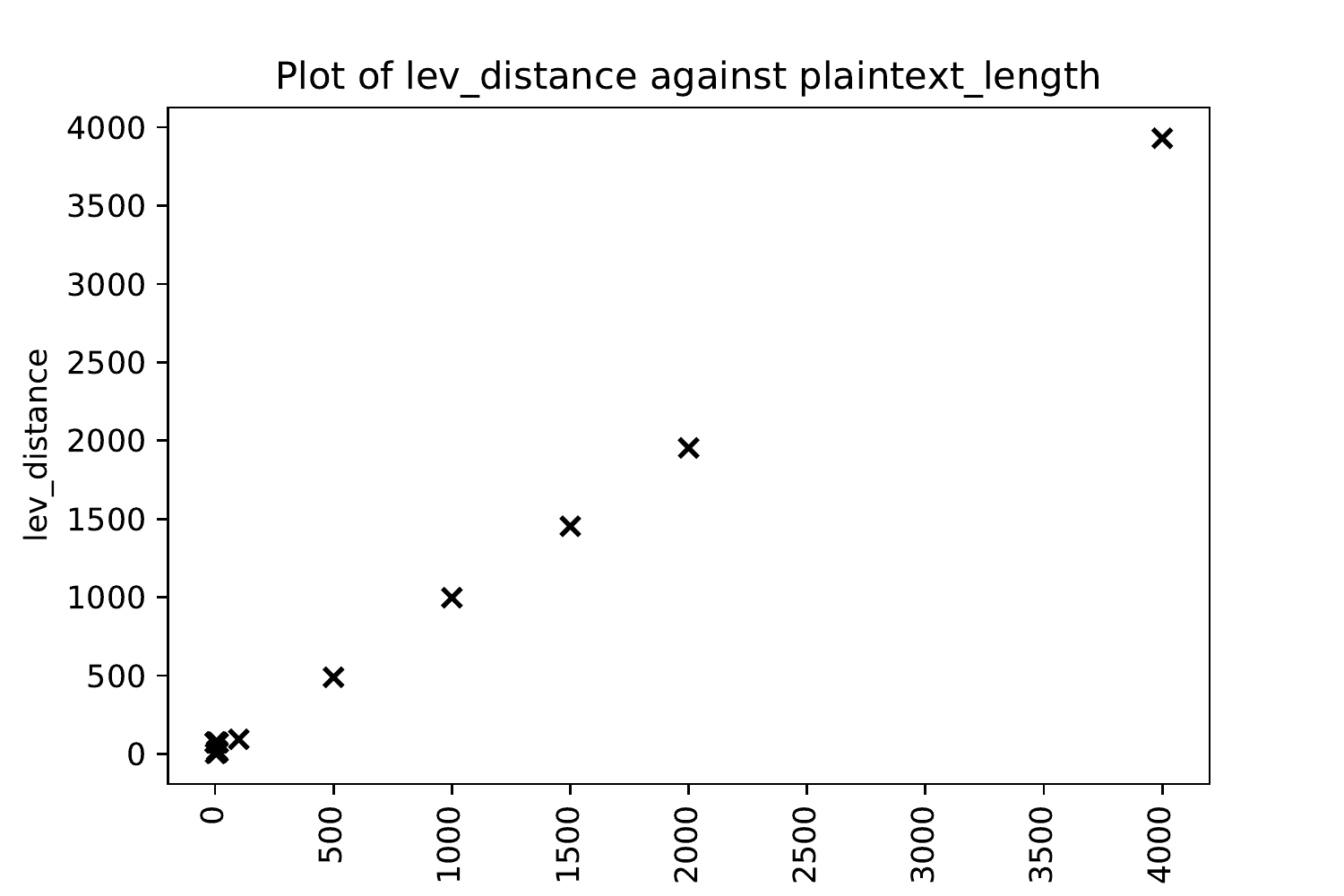}
  \caption{Plot of similarity metric against plaintext length}
\end{figure}

Beyond execution cost, this experiment yielded additional information regarding the properties of this obfuscation model. 

Figure 6 adds onto prior Stealth results in section 4.1 to reveal that the larger the code base, the greater the dissimilarity between the obfuscated code and the original code base.

Plotting ciphertext character variation and ciphertext length against plaintext length reveals:  (i) the character variation is widely distributed regardless of the length of the plaintext input, which further supports the notion of randomness of ciphertext generation, as the ciphertext is based purely on the randomness in the model weight generation; (ii) the ciphertext length is kept low (on average 72 character length) regardless of the plaintext length, which reduces obfuscated code storage requirements. The notion that the cipher generation algorithm produces short random ciphertexts further implies it would be difficult for malicious actors to reverse-engineer the ciphertext by setting random weights or training a model without a known output text. For the model used in the experiment, the model file consists of 3 layers of arrays: the main layer contains 8 arrays, of which each array has an array-length as represented in [39, 256, 1024, 72, 256, 1024, 256, 72], of which each sub-array (except the third and sixth array) contains an array of 1024 values. One would have to randomly generate 32-bit floats for 975,872 values, at least to a proximate range to the actual values before they can generate readable de-obfuscated code, but even then they would need to further generate values to the 8th decimal place if they intend to obtain scrambled text without any lapse in meaning. 

\begin{figure}[h!]
  \centering
  \includegraphics[scale=0.5]{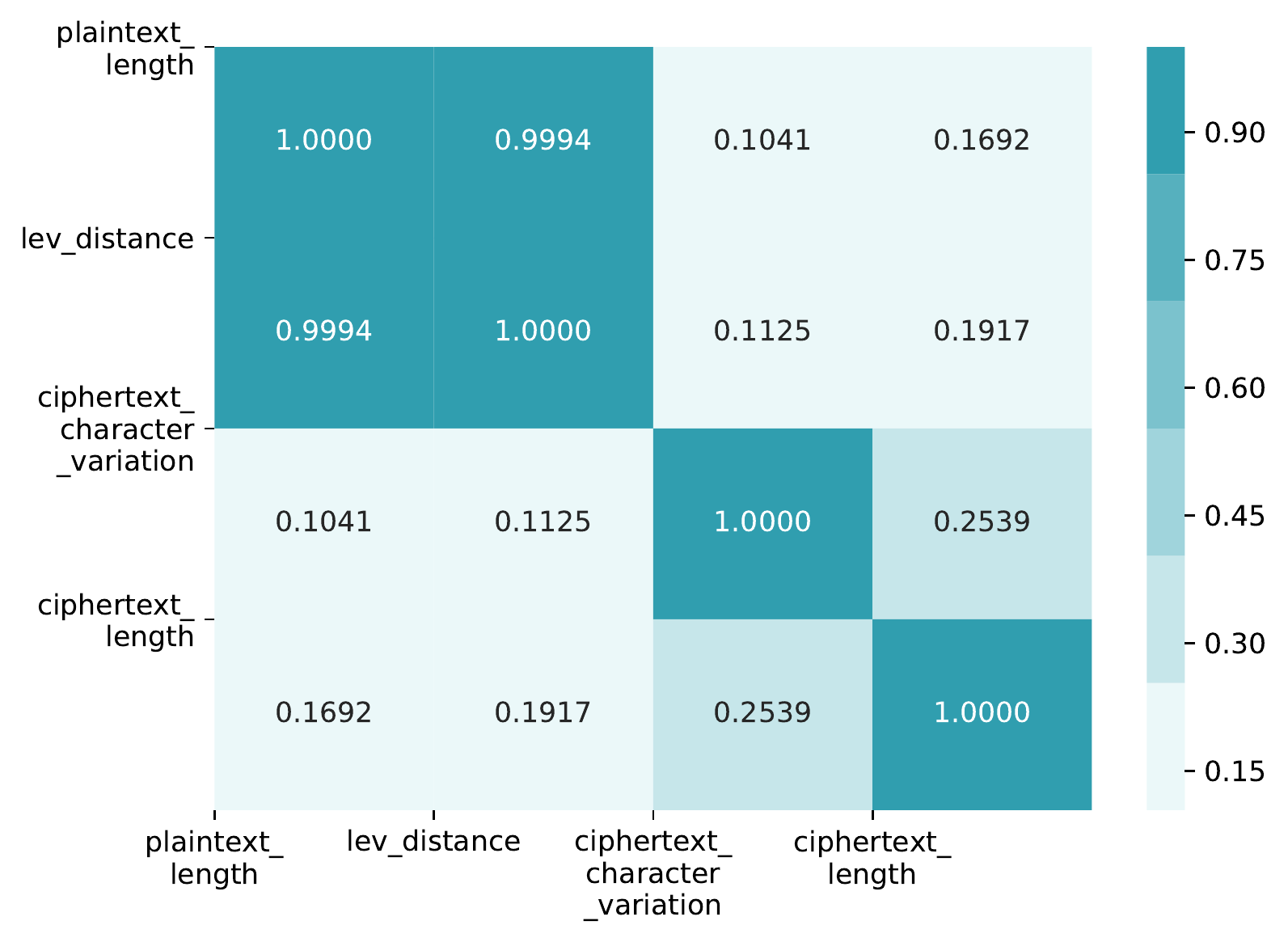}
  \caption{Correlation matrix of properties}
\end{figure}

The correlation matrix in Figure 7 summarizes the relationships between the properties tested in the experiment. 

\section{Conclusion}
This paper presents a novel obfuscation methodology using RNN encoder-decoder models to generate ciphertext from source code and generating and utilizing model weights as keys. Compared to current obfuscation methods, it is at least on par in terms of stealth and is expected to outperform for larger code bases in terms of obscurity and readability, and though key generation may take a significant amount of time for larger code bases or require more computational resources, it would be less time-intensive than to manually obfuscate the source code. This would be a good use case application for services that have confidential source code in plaintext but would prefer ciphertext yet require the ability to execute.

%
%

\end{document}